\documentclass[useAMS,usenatbib]{mn2e}
\usepackage{amsmath}
\usepackage{graphicx}
\usepackage{epstopdf}

\title{Grid Noise in Moving Mesh Codes: Fixing The Volume Inconsistency Problem}
\author{Elad Steinberg$^1$\thanks{E-mail: elad.steinberg@mail.huji.ac.il}, Almog Yalinewich$^1$ and Re'em Sari$^1$\\
$^1$ Racah Institute of Physics, Hebrew University, Jerusalem 91904, Israel}

\begin{document}

\maketitle

\input{aas_macros.sty}

\begin{abstract}
Current Voronoi based moving mesh hydro codes suffer from ``grid noise". We identify the cause of this noise as the volume inconsistency error, where the volume that is transferred between cells is inconsistent with the hydrodynamical calculations. As a result, the codes do not achieve second order convergence.
In this paper we describe how a simple fix allows Voronoi based moving mesh codes to attain second order convergence. The fix is based on the understanding that the volume exchanged between cells should be consistent with the hydrodynamical calculations. We benchmark our fix with three test problems and show that it can significantly improve the computational accuracy. We also examine the effect of initial mesh initialization and present an improved model for the Green-Gauss based gradient estimator.
\end{abstract}


\section{Introduction}
The use of moving mesh hydro codes, and in particular Voronoi based, in astrophysics are ever increasing. These codes offer the ability to accurately capture shocks, diminish diffusion errors and preserve discontinuities extremely well \citep{arepo,TESS,RICH}. An added benefit is the complete freedom to determine the location of the computational cells, allowing for high resolution in the areas of interest.

Unfortunately, current Voronoi based moving mesh codes exhibit grid noise \citep{BS12,MOCZ14,Hopkins14,Paul15}. This noise manifests itself in errors with wavelengths of order the cell size. The magnitude of the error is largest where the topological changes between cells is largest. Until recently, the exact nature of this noise was not well understood \citep{RICH}. Several authors have described various heuristic fixes for it. \cite{Paul15} have described a method that smooths the velocity that is assigned to the mesh points hence lowering the velocity difference between neighboring mesh points. This causes topological changes between neighboring mesh points to be smoother as well. \cite{MOCZ15} proposed a new method to regularize the mesh by successive Llyod iterations. This has the benefit of making the Voronoi cells ``round" even in regions where the topological changes might be sharp without it. Both suggested fixes indeed reduce the grid noise but do not eliminate it.

In addition to the problem of grid noise, it has been noted by \cite{PAKMOR15} that the method proposed by \cite{arepo} of estimating the gradient of a cell based on the Green-Gauss theorem does not converge. We suggest a slight variation to this method that allows the gradient estimate to converge.

The paper is organized as follows.
In Section \ref{sec:GG} we derive an improved Green-Gauss based gradient estimator. We describe our method of fixing the area inconsistency problem in Section \ref{sec:fix}. The relation to recent works regarding reduction of mesh noise is discussed in Section \ref{sec:recent}. Several examples illustrating how our fix improves the code are described in Section \ref{sec:res}.
\section{Improved Green-Gauss Gradient Estimate}
\label{sec:GG}
\subsection{Prerequisite for Second Order Convergence}
In order to achieve second order convergence in Godunov type schemes, the primitive variables, which are the input for the Riemann solver, must be linearly extrapolated from the cell center to the edges where the flux is calculated. Second order convergence can only be achieved if the error in the gradient estimation decreases at least as one over the resolution. 
\cite{PAKMOR15} have shown that the gradient estimation based on the Green-Gauss theorem that is currently used in Voronoi based moving mesh codes does not converge with resolution in general, but only for centrodial meshes. In order to achieve second order convergence in AREPO, they use a linear least squares fit to find the gradient. In the following section we show how a simple change enables the Green-Gauss gradient estimate to converge and result in second order convergence for the hydro scheme.
\subsection{Improving the Green-Gauss Gradient Estimate}
The original Green-Gauss based gradient estimate assumes that the information about the primitive variables are known at the mesh generating points. However, since the primitive variables are volume averaged quantities, their location is at the cell's center of mass. This gives rise to a relative error in the estimation of the gradient of the order $|\vec{s}-\vec{r}|/\sqrt{A}$, where $\vec{s}$ is the cell's center of mass, $\vec{r}$ is the location of the mesh generating points and $A$ is the cell's volume (area in 2D). In the following we derive a Green-Gauss based estimate that assumes, as is more accurate, that the primitive variables are known at the cell's center of mass.

If we label the gradient of a cell by $\vec{b}$, then according to the Green-Gauss theorem up to zeroth order in $\sqrt{A}$ we can approximate $\vec{b}$ to be:
\begin{equation}
\vec{b}\approx\sum_j\phi (\vec{f}_j)\vec{L}_j/A+\mathcal{O}(\sqrt{A})\label{eq:first}
\end{equation}
where $\phi$ is the quantity that we are calculating the gradient for, $\vec{f}_j$ is the middle of the j-th edge, $\vec{L}_j$ is a vector with magnitude equal to the length of the edge and pointing outward from the cell (normal to the edge) and the summation is performed over all of the cell's neighbors (i.e. over all of the edges).

The value of $\phi (\vec{f}_j)$ can be approximated as
\begin{equation}
\phi (\vec{f}_j)\approx \phi + \vec{b}\cdot(\vec{f}_j-\vec{s})+\mathcal{O}(A)
\end{equation}
where $\phi$ is the volume averaged quantity.
This can be simplified by defining $\vec{c}_j\equiv\vec{f}_j-(\vec{s}+\vec{s}_j)/2$. This is similar to what is defined in \cite{arepo} but with the replacement of the mesh generating points with the center of masses.

Equation \ref{eq:first} now becomes
\begin{equation}
\vec{b}\approx\sum_j[\phi +\vec{b}\cdot(\vec{c}_j+(\vec{s}_j-\vec{s})/2)]\vec{L}_j/A+\mathcal{O}(\sqrt{A})
\end{equation}
Approximating the value of the neighbor as $\phi_j/2\approx\phi/2+\vec{b}\cdot(\vec{s}_j-\vec{s})/2$, gives 
\begin{equation}
\left(\mathbf{I}-\sum_j\vec{c}_j \otimes \vec{L}_j/A\right)\vec{b}=\sum_j\vec{L}_j(\phi+\phi_j)/2A+\mathcal{O}(\sqrt{A}) 
\end{equation}
where $\otimes$ denotes outer product. 
The matrix $\mathbf{D}$ is defined as
\begin{equation}
D_{l,k}=\sum_j c_{j,k}L_{j,l}/A
\end{equation}
where $l,k$ are the coordinates ($x,y$). This gives us
\begin{equation}
(\mathbf{I}-\mathbf{D})\vec{b}\approx\sum_j\vec{L}_j(\phi+\phi_j)/2A+\mathcal{O}(\sqrt{A}) 
\end{equation}
Inverting the left hand matrix, $\mathbf{E}=(\mathbf{I}-\mathbf{D})^{-1}$, gives us
\begin{equation}
\vec{b}=\mathbf{E}\cdot\sum_j\vec{L}_j(\phi+\phi_j)/2A+\mathcal{O}(\sqrt{A})
\end{equation}
which converges as one over the resolution as needed. 
This derivation is trivially extended to 3D by the substitution of the area of the cell with its volume and by replacing $\vec{L}$ with a vector whose magnitude is the area of the relevant Voronoi face and is pointing normal to the face.

\begin{figure}
	\centering
	\includegraphics[width=0.52\textwidth]{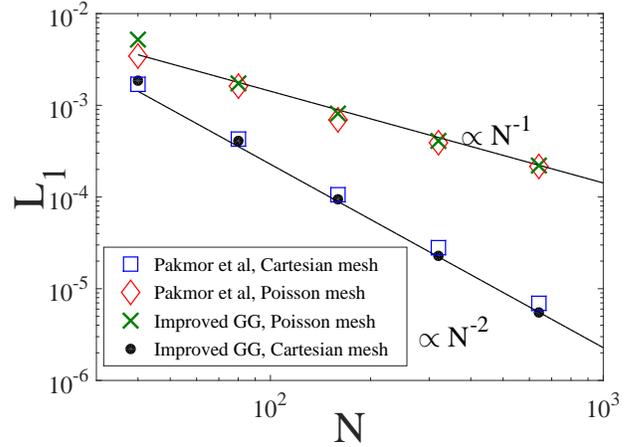}
	\caption{$L_1$ norm of the gradient estimate of the density field for different types of meshes for the initial state of the Yee vortex at t = 0.}
	\label{fig:grad}
\end{figure}

Our end result resembles the one shown by \cite{arepo} with the replacement  
\begin{equation}
\mathbf{D}\cdot\vec{b}\Rightarrow \sum_j L_j(\phi_j-\phi)\frac{\vec{f}_j-\frac{\vec{r}+\vec{r}_j}{2}}{A|\vec{r}-\vec{r}_j|}.
\end{equation}
Transforming the old Green-Gauss gradient estimate to the improved one requires very little coding.
In contrast to \cite{arepo}, where the gradient is calculated relative to the mesh generating point but is used to interpolate from the center of mass, we calculate the gradient and interpolate relative to the center of mass.
In figure \ref{fig:grad} we present the results of our improved gradient estimate compared to the least squares method of \cite{PAKMOR15}. The gradient is estimated for the Yee vortex problem (described in detail in section \ref{sec:yee}) and compared with the analytical result. The gradient estimate is calculated once using a Cartesian mesh (which is centrodial) and once for a mesh whose points are Poisson sampled based on the density distribution. Our improved gradient estimate converges linearly as one over the resolution for the Poisson sampled mesh and quadratically for the Cartesian mesh. Our method gives comparable results compared to the least squares method of \cite{PAKMOR15}.

\section{Fixing the Area Inconsistency Problem}
\label{sec:fix}
\subsection{The Area Inconsistency Problem}
As stated in\cite{RICH}, there is an inconsistency between the hydrodynamical calculation and the actual change in the volume (area in 2D, as we assume from here onwards) of the cell in Voronoi based moving mesh codes. During each timestep, the length and the velocity of the edge is assumed constant. This implicitly assumes that the edge sweeps over an area
\begin{equation}
\Delta A_{flux}=v_nL\Delta t
\end{equation}
where $v_n$ is the normal component of the edge's velocity and $L$ is the edge's length. 
However, the actual change in the cell's area is not necessary equal to $\Delta A_{flux}$ and can be quite different.
When all the mesh generating points have the same velocity, $\Delta A_{flux}$ is equal to the actual change in the cell's area, $\Delta A_{real}$. 

Simple dimensional analysis shows that
\begin{equation}
\frac{\Delta A_{flux}-\Delta A_{real}}{A}\propto\frac{\Delta v R\Delta t}{A}\propto\frac{\Delta v}{c_s}
\end{equation}	
where $\Delta v$ is the order of magnitude of the velocity difference between neighboring mesh generating points, $c_s$ is the speed of sound, $A$ is the cell's area and $R=\sqrt{A}$. If the flow is smooth and the velocities of the mesh generating points are moved Lagrangianly (i.e. with the local fluid velocity), then the velocity difference between neighboring mesh generating points scales inversely with the resolution. The relative error then scales as
\begin{equation}
\frac{\Delta A_{flux}-\Delta A_{real}}{A}\propto\frac{\Delta v}{C_s}.
\end{equation}
where in a smooth flow $\Delta v\propto R$. This shows that for a smooth flow with the mesh generating points moving Lagrangianly, the error decreases inversely with the resolution.
However, for stability reasons it is required that the Voronoi cells be rather ``round". Since this mesh regularization is achieved by assigning the mesh generating points a velocity that is not sampled from a smooth field, the velocity difference between neighboring mesh generating points can be of order the sound speed. This results in a relative area difference that is resolution independent, thus preventing Voronoi based codes from properly converging.

\begin{figure}
\centering
\includegraphics[width=0.52\textwidth]{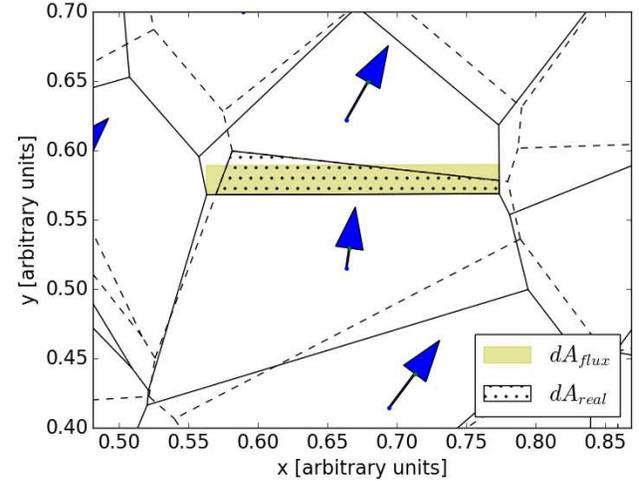}
\caption{Example of the difference between the assumed area change in the hydrodynamical calculation and the actual area change between cells. The assumed area change, $\Delta A_{flux}$ is shown in the colored region while the actual area exchanged is shown by the spotted area. Arrows denote the velocities of the mesh generating points, the black solid line denotes the old mesh and the dashed line the new mesh.}
\label{fig:mesh}
\end{figure}

Figure \ref{fig:mesh} shows an example of the assumed $\Delta A_{flux}$ and the actual area exchanged, $\Delta A_{real}$.

Since $\Delta t\sim R/c_s$, higher order time integration schemes cannot change the scaling of the error with the resolution.
Our fix, described below, allows the code to properly converge with first order accuracy. In order for the code to converge with second order accuracy, this fix must be applied in tandem with standard second order methods (e.g. gradient extrapolation of the primitive variables coupled with a Runge-Kutta time integration scheme).

The greater the deformation, the higher is the error. This explains why in post shock regions, where cells get suddenly compressed, or near shear boundaries, the mesh noise is the largest.

\subsection{The Fix}
In order to fix the above problem, one must compensate for the discrepancy between $\Delta  A_{flux}$ and the actual change in the cell's area, $\Delta A_{real}$.

\cite{ReALEPaper} solved a related problem in their ALE code ReALE. In ReALE the vertices of the Voronoi edges are moved in a Lagrangian fashion and each time step the distorted Voronoi cells are mapped onto a new Voronoi mesh. This mapping is done by finding the polygon clippings of the distorted cell with the new Voronoi mesh. 

We propose a similar scheme, and have implemented it in our code RICH. The outline of the method is as follows:
\begin{enumerate}
    \item Calculate the fluxes.
\item Move the mesh points and construct a new Voronoi mesh.
\item Find the intersections between new Voronoi cells and the old mesh.
\item Calculate for each intersection, a modified area, $\delta A$, which is the area of the intersection between the two polygons minus the relevant $\Delta A_{flux}$.
\item Denoting $U$ the conserved variables per unit area; transfer an amount of $\delta A\cdot \tilde{U}$ of conserved variables between the cell that has lost area to the cell that gained area , where $\tilde{U}$ is taken as the average between the cells if there is less than a factor of two difference between the cells and solely from the donor cell if the factor is larger.
\end{enumerate}
Preliminary results show that this fix enables second order convergence but at the expense of a heavy computational cost. Calculating the polygon intersections takes about 5 times as much time as running the rest of the code, so it is not a reasonable approach.

Based on \cite{ReALE} we propose an alternative scheme that approximates the above. We modify steps (iii) and (iv) as follows:
\begin{enumerate}
\setcounter{enumi}{2}
\item Find for each edge in the old mesh its corresponding edge in the new mesh. This is achieved by the requirement that the edges have the same neighbors.
\item $\delta A$ is now calculated as the difference between the quadrangular defined by the two edges and $\Delta A_{flux}$.
\end{enumerate}

Since not all of the edges in the old tessellation have corresponding edges in the new tessellation, special care is required to deal with those edges. We discuss this in the appendix.
The advantage of this approximate scheme, is that the run time for the fix is only about $15\%$ of the code's total run time.
\section{Relation to Recent Works}
\label{sec:recent}
Both \cite{Paul15} and \cite{PAKMOR15} have addressed similar issues in their works. 

\cite{Paul15} correctly related the shear velocity between neighboring mesh generating points as contributing to grid noise. Large shear velocity typically induce a large area inconsistency error since the shear velocity is not taken into account during the velocity of a cell's edge calculation. Their proposed fix to smooth out the velocities assigned to the mesh generating points in order to minimize the shear velocity reduces the error. However, as we will show, this smoothing does not eliminate the error since it only reduces the shear velocity and does not eliminate it. 

\cite{PAKMOR15} have presented a method for second order time integration in moving mesh codes as well as correctly pointing out the error in the Green-Gauss based gradient estimate of \cite{arepo} and fixing it. Their method has the benefit of requiring only one mesh construction per time step and thus shortens the run time. While their method achieved second order convergence for a mesh that captures the symmetry of the problem, it failed to achieve second order convergence for a general mesh ordering.
\cite{MOCZ15} proposed a method to give the mesh generating points a correction velocity based on trying to predict where the centroid of the Voronoi cell will be. As we show in section \ref{sec:res}, this method lowers significantly the area inconsistency error but fails to get rid of it and at high resolution fails to achieve second order convergence.

Both of these works help improve the accuracy of Voronoi based moving mesh codes and go hand in hand with our work. A second order time integration scheme coupled with a converging gradient estimator as well as applying our fix is required to achieve second order convergence. 

\section{Results}
\label{sec:res}
In this section we use the RICH code \citep{RICH} to compare results with and without our fix for the area inconsistency problem. All of the tests are run with a CFL number of 0.6 and use a second order Runge-Kutta time integration scheme (midpoint method).
\subsection{Pure Advection}
A simple problem which all codes should handle well is a pure advection problem in 2D. The initial setup for this problem is
\begin{equation}
\begin{split}
\rho&=1\\
P&=1\\
v_x&=1\\
v_y&=0
\end{split}
\end{equation}
with periodic boundary conditions on the unit square. This initial profile should remain constant for all times. 
Naively, in this problem there should be no area inconsistency error since in our Lagrangian scheme all of the cells have the same velocity. However, the mesh points have an additional velocity that tries to make the Voronoi cells ``rounder". This additional velocity causes neighboring cells to have different velocities and gives rise to the area inconsistency problem.
We run this setup twice up to $t=1$. The first run is without the area inconsistency fix and the second is with it. For both runs the mesh generating points are sampled from a uniform random distribution and then relaxed with 20 Llyod iterations. 
\begin{figure}
\centering
\includegraphics[width=0.52\textwidth]{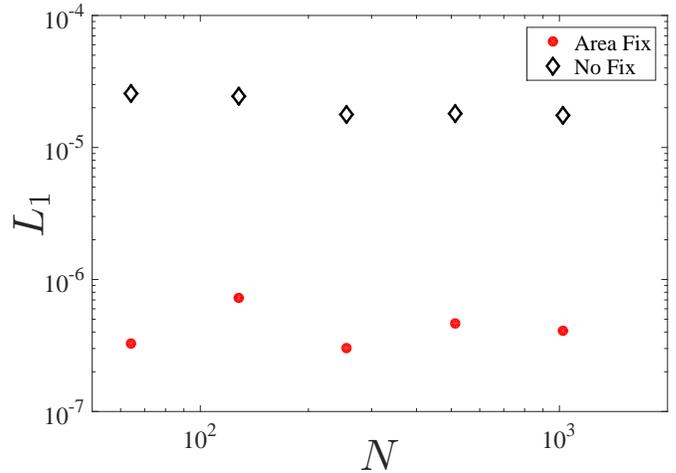}
\caption{The $L_1$ error norm of the density as a function of linear resolution for the pure advection test.}
\label{fig:advection}
\end{figure}
In figure \ref{fig:advection} we plot the $L_1$ norm of the error in the density defined as
\begin{equation}
L_1=\frac{\sum_i|\rho_i-1|A_i}{\sum_i A_i}.
\end{equation}
Without the area inconsistency fix, the error is resolution independent in agreement with the prediction and is about $\sim 2\cdot 10^{-5}$. 
Since the velocity and pressure are uniform, the mass is advected along with the flow and traces the change in the area of a cell. Hence, the error in the density can only arise from the area inconsistency.

When applying the fix, we once again see an error that is roughly resolution independent (even though we used a second order time integration scheme) but is about a factor $40$ smaller than the previous scenario. This error arises from the few rare cases which we do not deal with (e.g. a cell changing two adjacent neighbors). 
On a side note, when running this test with the polygon clipping method, the $L_1$ error is comparable to machine precision.
Overall, our fix improves the performance of RICH and reduces the $L_1$ error by a factor $~40$ for this test.
\subsection{Yee Vortex}
\label{sec:yee}
The Yee vortex test \citep{Yee2000} is a good test problem to see how our fix handles smooth flows. In this test, isentropic vortices that balance the centrifugal force and the pressure gradient are evolved. The setup is
\begin{eqnarray}
\rho(r) & = & \left(T_{inf}-\frac{(\gamma-1)\beta^2}{8\gamma\pi^2}e^{1-r^2}\right)^{1/(\gamma-1)}\\
P(r) & = & \rho^\gamma/\gamma\\
v_x(r)& = & -y\frac{\beta}{2\pi}e^{(1-r^2)/2}\\
v_y(r)& = & x\frac{\beta}{2\pi}e^{(1-r^2)/2}
\end{eqnarray}
and the parameters are set to be $T_{inf}=1,\;\beta=5$ and $\gamma=1.4$.

We run this test for three different methods for setting the velocity of the mesh generating points. The first is the one suggested by \cite{MOCZ15}, which we call centroid motion, the second is the one described in \cite{arepo} and \cite{RICH}, which we call standard motion and the third is the one proposed by \cite{Paul15}, which we call smooth motion. For each one of the three methods we preform a run with and without our suggested fix. For all of the runs the initial mesh is drawn from a uniform random distribution and relaxed with 20 Llyod iterations.
\begin{figure}
	\centering
	\includegraphics[width=0.52\textwidth]{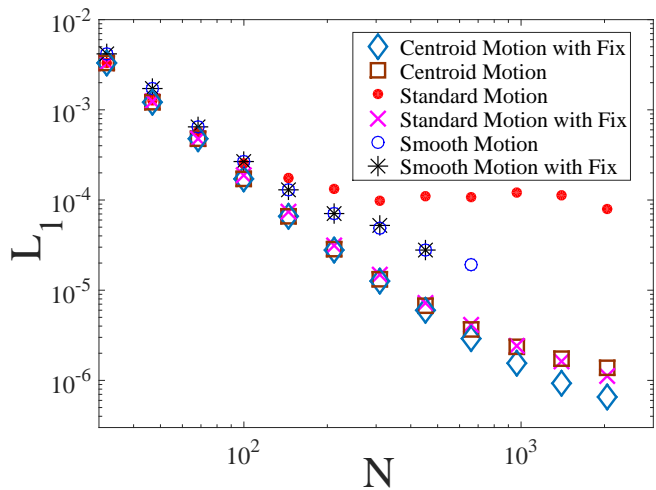}
	\caption{The $L_1$ error norm of the density as a function of linear resolution for the Yee vortex test.}
	\label{fig:Yee}
\end{figure}

In figure \ref{fig:Yee} we show the $L_1$ error at time $t=10$ for all of the runs.	 
At low resolutions all of the runs give comparable results. Once the $L_1$ error norm of the density is approximately $\sim10^{-4}$ the standard motion without our fix starts to level off. 

The runs with the smooth motion suffer less from the area inconsistency error than the standard motion since the velocity difference between mesh generating points is lower. However, the high resolution runs failed to complete due to cells with very high aspect ratio which give rise to a large error in the hydrodynamic scheme. This method is very unstable without adding an additional velocty term to make the cells ``rounder". 

Moving the mesh generating points with centroid motion gives a significantly better result than the other two motions (without applying the fix), but it starts to level off when $L_1\sim10^{-6}$. This result rises naturally from the area inconsistency problem. The correction term in the velocity of the mesh generating point is a factor of $d/R$ smaller for centroid motion relative to the standard motion, where $d$ is the distance between the cell's center of mass and the mesh generating point and $R$ is the cell's width. \cite{MOCZ15} have shown that this factor is typically $\sim0.01$, thus we expect the area inconsistency error to be about two orders of magnitude smaller for this scheme. 
It is worth mentioning that if the correction velocity in the standard motion is changed from $c_s$ to $c_sd/R$ then the original point steering method and the one introduced in \cite{MOCZ15} give comparable results at all resolutions.
When our fix is applied the $L_1$ error norm starts to level off a factor of a few less then for the centroid motion due to the rare cases which we don't account for in our fix, in agreement with the results of the previous section.

\subsection{Noh Problem}
The Noh problem \citep{Noh1987} checks how the code handles strong
shocks and highly supersonic flow. The setup for the test is a uniform
density $\rho=1$, small uniform pressure $P=10^{-6}$ and uniform
radial inflow velocity $v=1$ while the adiabatic index is set to $\gamma=5/3$.

Our computational domain is ${\left[-1,1\right]}^{2}$ and we use $10^4$ mesh generating points, randomly distributed
across the domain and relaxed with 10 Llyod iterations, and the boundary conditions are dictated from the analytic solution. We split
cells once their volume increases above 150\% of their initial value
and remove them when their volume drops below 25\% of their initial
value in order to prevent pile up of cells at the center and too large cells at the boundaries. We run the test once without the fix and once with it.
\begin{figure}
\centering
\includegraphics[width=0.49\textwidth]{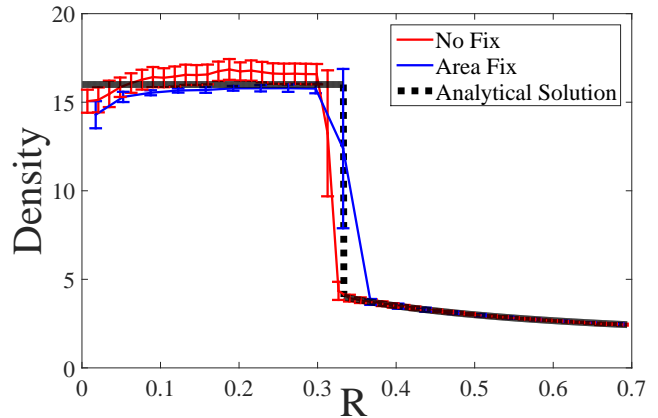}
\caption{The density as a function of radius of the Noh test. The error bars are calculated by taking the $1\sigma$ deviation of cells in a given radial bin.}
\label{fig:Noh}
\end{figure}
Figure \ref{fig:Noh} shows the density as a function for the radius for both runs as well as the analytical solution overlaid in the black line. The most pronounced difference between the runs is in the post shock area where our fix suffers less noise. This is a direct result of how the area inconsistency error produces noise behind shock fronts.
\section{Conclusions}
We have presented a method to fix the area consistency error that has been labeled previously as ``grid noise". This noise prevents Voronoi based moving mesh hydro codes from converging. Our tests show that this fix indeed allows our code to achieve second order convergence, and greatly improves its accuracy. This fix is not computationally expensive, is easy to implement and can be extended to non-Voronoi based moving mesh codes that also suffer from the area inconsistency problem. 

Our proposed fix goes hand in hand with other recent works regarding methods to reduce mesh noise by correcting the velocities of the mesh generating points. Improving the methods of assigning velocities to the mesh generating points can reduce both the hydrodynamical errors as well as reduce the area inconsistency error. Our fix greatly reduce the area inconsistency error and is best used in tandem with a scheme that moves the mesh generating points in such a way as to maintain ``roundish" cells (but doing this with a low velocity compared to the sound sped), while keeping the mesh motion as close to Lagrangian as possible.
If our fix is not applied, it is crucial to keep the velocity difference between mesh generating points as small as possible (including their correction terms) or the area inconsistency error can be high.

\section*{Acknowledgments}
We would like to thank Paul Duffell for helpful comments and fruitful discussions. This research is supported in part by  ISF, ISA, iCORE grants and a Packard Fellowship.

\bibliographystyle{mn2e}
\bibliography{AreaFix}

\appendix
\label{sec:flip}
\section{Flipped Edge}
The most common scenario when an edge does not have a corresponding edge in the new tessellation is an ``edge flip" scenario.
Other cases, such as a cell changing two adjacent neighbors we neglect and apply no fix. These cases are rare and do not impact the convergence as was seen. 
An ``edge flip" occurs when there is an edge, $E_1$, in the old mesh and an edge, $E_2$, in the new mesh with the following relation. $E_1$ has neighbors $n_0$ and $n_1$, and both $n_0$ and $n_1$ have two mutual neighbors $n_2$ and $n_3$. $E_2$ has neighbors $n_2$ and $n_3$, and both $n_2$ and $n_3$ have two mutual neighbors, $n_0$ and $n_1$. 
\begin{figure}
\centering
\includegraphics[width=0.52\textwidth]{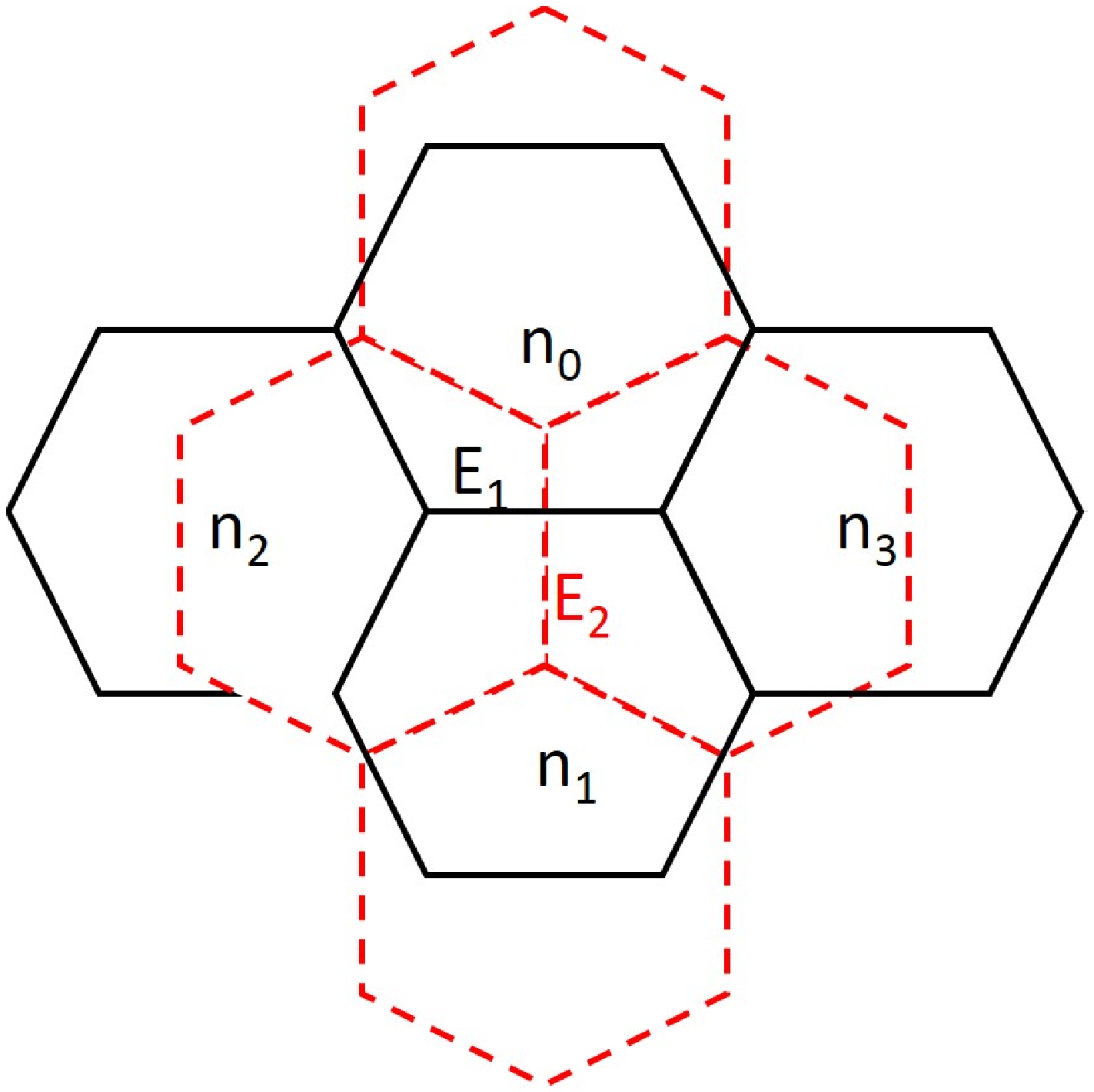}
\caption{An example of how edge $E_1$ flipped into edge $E_2$. The solid black tessellation is the tessellation at the beginning of the time step and the dashed red tessellation is at the end of the time step.}
\label{fig:flip}
\end{figure}
\begin{figure}
\centering
\includegraphics[width=0.52\textwidth]{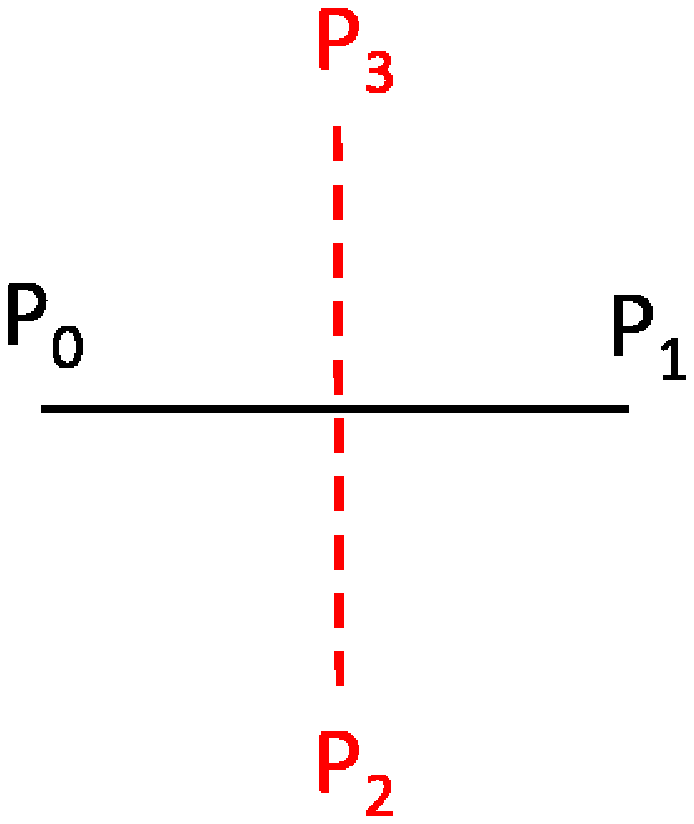}
\caption{The four vertices that form the flipped edges. Edge $E_1$ is composed of vertices $P_0$ and $P_1$ while edge $E_2$ is composed of vertices $P_2$ and $P_3$.}
\label{fig:flip2}
\end{figure}
Figures \ref{fig:flip} and \ref{fig:flip2} shows an illustration of such an occurrence when the 4 cells exchange neighbors among themselves. When this occurs, each one of the 4 cells calculates the area it has lost or gained. This is done by computing the area of the triangle composed of the relevant edge and the cell's boundary. For example, in figure \ref{fig:flip}, cells $n_0$ and $n_1$ lose area while cells $n_2$ and $n_3$ gain it. Cell $n_0$ lost the area from the triangle formed by the points $P_0P_1P_3$, cell $n_1$ lost the area from the triangle formed by the points $P_2P_1P_0$, while cell $n_2$ gained the area from the triangle formed by the points $P_2P_3P_0$ and cell $n_3$ gained the area from the triangle formed by the points $P_1P_3P_2$.

We calculate $\Delta A_{flux}$ only for the edge in the old mesh since the new edge has no flux calculated for it.
If a cell has a negative $\delta A$, it donates $-\delta A\cdot U$, while if a cell has a positive $\delta A$, it gains $\delta A\cdot U_{tot}$ where 
\begin{eqnarray}
U_{tot}&=&\sum_i -\delta A_i\cdot U_i/A_{tot}\\
A_{tot}&=&\sum_i -\delta A_i
\end{eqnarray}
and both summations are taken only over cells with negative $\delta A$.
The creation and destruction of new edges almost always occurs via edge flips. The only scenarios where there are edges created/destroyed without edge flips is the degenerate case of an edge shrinking exactly to zero length or via interactions with the computational domain. The former scenario is extremely rare while the latter produces no area inconsistency error since the domain walls do not move.

\end{document}